\documentclass[useAMS,usenatbib]{mn2e}

\usepackage{graphicx}

\bibpunct{(}{)}{;}{a}{}{ }

\begin{document}
\title[Black hole binary in M31 globular cluster]{A second black hole candidate in a M31 globular cluster is identified with XMM-Newton}
\author[R. Barnard, and U. Kolb]{R. Barnard$^1$ and U. Kolb$^1$\\
$^{1}$Department of Physics and Astronomy, The Open University, Walton Hall, Milton Keynes, MK7 6AA, UK}
\date{}

\pagerange{\pageref{firstpage}--\pageref{lastpage}} \pubyear{2009}

\maketitle

\label{firstpage}

\begin{abstract}
We use arguments developed in previous work to identify a second black hole candidate associated with a M31 globular cluster, Bo 144, on the basis of X-ray spectral and timing properties.  The 2002 XMM-Newton observation of the associated X-ray source (hereafter XBo 144) revealed  behaviour that is common to all low-mass X-ray binaries (LMXBs) in the low-hard state. Studies have shown that neutron star LMXBs exhibit this behaviour at 0.01--1000 keV luminosities $\le$10\% of the Eddington limit ($L_{\rm Edd}$).  However, the unabsorbed 0.3--10 keV of XBo 144 luminosity was $\sim$0.30 $L_{\rm Edd}$ for a 1.4 M$_{\odot}$ neutron star, and the 0.01--1000 keV luminosity is expected to be  $\ga$3--7 times higher. We therefore identify XBo 144 as a black hole candidate. Furthermore, it is the second black hole candidate to be consistent with formation via tidal capture of a main sequence donor in a GC; such systems were previously thought non-existent, because the donor was thought to be disrupted during the capture process.
\end{abstract}


\section{Introduction}

In \citet{bsh08}, we identified the first genuine black hole candidate using our X-ray classification  method, which was developed to the fullest in \citet{bsh08}. The X-ray source associated with the globular cluster  Bo 45 (hereafter known as XBo 45) exhibited  behaviour associated  with all low mass X-ray binaries (LMXBs) in the low state: hard power law emission, and high r.m.s variability \citep[see e.g.][]{vdk94, vdk95}. A  recent survey of Galactic neutron star LMXBs has shown that the low state is observed at 0.01--1000 keV luminosities $\la$0.1 $L_{\rm Edd}$ for systems that trace a diagonal transition in colour-colour space from low state to high state,  and $\la$0.02 $L_{\rm Edd}$ for systems that exhibit a vertical transition \citep{glad07}; $L_{\rm Edd}$ is the Eddington luminosity.  However, XBo 45 exhibited this behaviour at a 0.3--10 keV luminosity $\sim$120 \% Eddington for a 1.4 M$_{\odot}$ neutron star, hence we identified it as a black hole candidate. 

XBo 45 is particularly interesting because there has been a distinct lack of black holes in globular clusters (GCs), and most black hole LMXBs are thought to be ejected from the cluster. The theoretical work of \citet{kal04} showed that one possible formation channel for black hole LMXBs is the tidal capture of a main sequence star; indeed, it is the most likely channel for dense clusters, such as those with collapsed cores.  \citet{kal04} predicted such systems to be persistently bright, and inferred  from the absence of such systems that the prospective donor is disrupted during capture. XBo 45 has appeared in all X-ray observations covering that region of sky, spanning $\sim$30 years, and hence is consistent with theoretical predictions for a system formed by tidal capture.

In this paper, we use the same arguments to identify  XBo 144 \citep[r2-5 in ][ $\alpha$ = 00:42:59.803 $\delta$ = 41:16:06.01]{K02}  as another  black hole candidate. Furthermore, XBo 144 is  located in a confirmed M31 GC \citep[see the revised Bologna Catalogue  V.3.5, March 2008;][]{gall04,gal05,gal06,gal07}.

 We present detailed analysis of the 2002 June XMM-Newton observation of XBo 144, the deepest of four exposures made between 2000 and 2002. Results from the other observations are consistent with this one.
In Section~\ref{obs} we describe the observations and data reduction, then present the results of our analysis in Section~\ref{res}. We discuss our findings in Section~\ref{discuss}, and draw conclusions in Section~\ref{conc}.

\section{Observations and data analysis}
\label{obs}

We rebuilt the data products for the 2002 June 26 XMM-Newton observation of the central region of M31  using version 7.1 of the Science Analysis Software suite (SAS). In order to screen for background flares, we followed the recommended procedure, finding a small flare near the start of the observation; this was removed. We then synchronised the pn and MOS lightcurves as described in \citet{bs07}.

 We extracted pn \citep{stru01}, and MOS \citep[MOS1 and MOS2][]{turn01} data from from a  circular source region with radius 20$''$, with a 20$''$ background region at a similar off-axis angle, with no point sources and on the same CCD chip. A larger background is desirable, but was  prevented by crowding.
Source and background lightcurves were constructed from the pn and MOS  data in 0.3--2.5, 2.5--10 and 0.3--10 keV bands. The lightcurves were background subtracted.  Source and background spectra were then created, along with corresponding response matrices and ancillary response files. The pn image yielded 5542 net source photons, while the data from the two MOS cameras were combined, yielding 4893 net source photons.

\section{Results}
\label{res}

\subsection{Spectral analysis}

The pn and combined MOS spectra were modeled simultaneously, using XSPEC ver 11.3. Each model consisted of an emission model, with line-of-sight absorption and a normalisation constant that accounts for differences in the pn and MOS callibrations, after setting the pn normalisation to 1. The  parameters of the emission models were forced to be  the same for the  pn and MOS spectra, but free to vary. Three emission models were initially chosen: blackbody (BB), bremsstrahlung (BR), and power law (PL). The blackbody and bremsstrahlung models were characterised by k$T$, where $T$ is the temperature and k is the Boltzmann constant, while power law emission is characterised by photon index, $\Gamma$. Absorption is expressed in terms of $N_{\rm H}$, the equivalent absorption by neutral hydrogen.

We first modelled the  XBo 144 spectra with single component emission models : BB, BR and PL. We found the pn and MOS spectra to be best fitted by the PL model, with $N_{\rm H}$ = 8.1$\pm$1.2$\times$10$^{20}$ atom cm$^{-2}$ and $\Gamma$ = 1.48$\pm$0.04; all spectral models are provided in Table~\ref{specmod}. The best fit model is as expected for a LMXB in its low state; however, we do not expect such behaviour at X-ray luminosities above $\sim$0.1 $L_{\rm Edd}$ in the 0.01--1000 keV band \citep{bsh08}. The 0.3--10 keV luminosity for our best fit PL model is 5.33$\pm$0.03$\times$10$^{37}$ erg s$^{-1}$, which is 0.29 $L_{\rm Edd}$ for a 1.4 $M_{\odot}$ neutron star, and 0.20 $L_{\rm Edd}$ for the most massive observed neutron star \citep[2.1 M$_{\odot}$, see e.g. ][]{nice05}. This is sufficently bright for XBo 144 to be a plausible black hole candidate.

\begin{figure}
\resizebox{\hsize}{!}{\includegraphics[angle=270,scale=0.6]{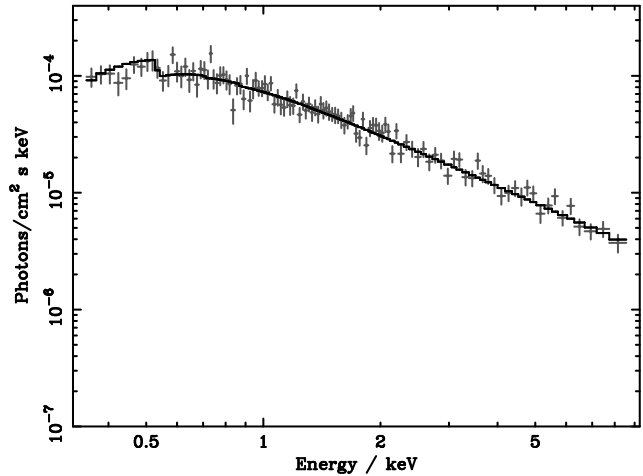}}
\caption{ Unfolded 0.3--10 keV pn spectrum of XBo 144 with best fit PL emission model.  }\label{plspec} 
\end{figure}

\begin{figure}
\resizebox{\hsize}{!}{\includegraphics[angle=270,scale=0.6]{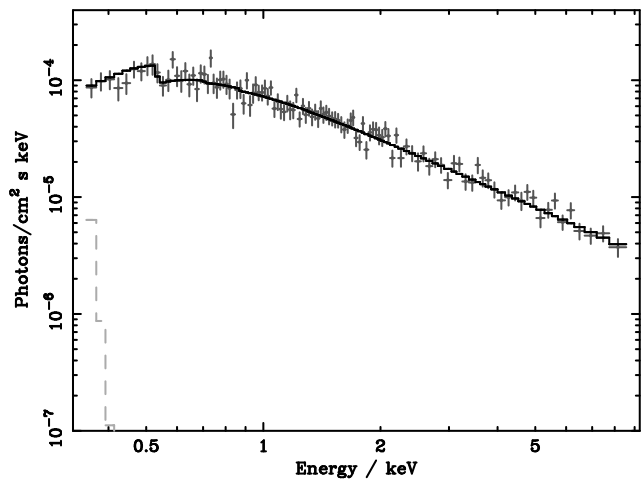}}
\caption{ Unfolded 0.3--10 keV pn spectrum of XBo 144 with best fit BB+PL emission model. The PL component is represented by a solid line, while the BB component is represented by a dashed line.  }\label{bbplspec} 
\end{figure}

We then modelled the XBo 144 spectra with a two component (BB+PL) emission model. The favoured blackbody temperature is 0.0082$\pm$0.0016 keV; this tells us that there is no blackbody component in the 0.3--10 keV band. In Figs.~\ref{plspec} and \ref{bbplspec}, we compare the PL and BB+PL emission models with the 0.3--10 keV pn spectrum, unfolded from the instrumental responses; we see that the blackbody component in Fig.~\ref{bbplspec} is set to have minimal contribution to the emission.

We also modelled the pn and combined MOS spectra separately with a PL emission model, to see if the instruments are in agreement. The best fit models are also shown in Table~\ref{specmod}; the pn and MOS spectra are in excellent agreement. We are therefore confident in our interpretation of the X-ray emission from XBo 144.

\begin{table*}
 \centering
  \caption{ Best fits to XBo 144 pn and MOS spectra with various emission models. Blackbody (BB), bremsstrahlung (BR) and power law (PL) emission models were used, all suffering line-of-sight absorption. For each model, we provide the absorption ($N_{\rm H}$ / 10$^{20}$ atom cm$^{-2}$), temperature (k$T$ / keV),  photon index ($\Gamma$), and MOS normalisation constant ($n_{\rm MOS}$). Numbers in parentheses indicate 90\% confidence limits. We then provide the $\chi^2$/dof. No uncertainties were calculated for the BB model, as $\chi^2$/dof = 8.5.}\label{specmod}
  \begin{tabular}{cccccc}
  \noalign{\smallskip}
  \hline
  \noalign{\smallskip}
  Model & $N_{\rm H}$   & k$T$ & $\Gamma$ & $n_{\rm MOS}$& $\chi^2$/dof\\
\noalign{\smallskip}
 \hline
\noalign{\smallskip}  
BB & 7  & 0.68 & $\dots$ & 1.07 & 1690/199\\
BR & 5.3(9) & 17(14) & $\dots$ & 1.11(4) & 193/198\\
PL & 8.1(12) & $\dots$ & 1.48(4) & 1.11(4) & 182/198\\
BB+PL & 8.7(12) & 0.0087(9) & 1.49(4) & 1.11(4) & 178/196\\ 
PL (pn) & 8.4(16) & $\dots$ & 1.46(6) & $\dots$ & 105/111\\
PL (MOS) & 7(2) & $\dots$ & 1.49(7) & $\dots$ & 74/85\\
\noalign{\smallskip}
\hline
\noalign{\smallskip}
\end{tabular}
\end{table*}

\subsection{Variability studies}
\begin{figure}
\resizebox{\hsize}{!}{\includegraphics[angle=270,scale=0.6]{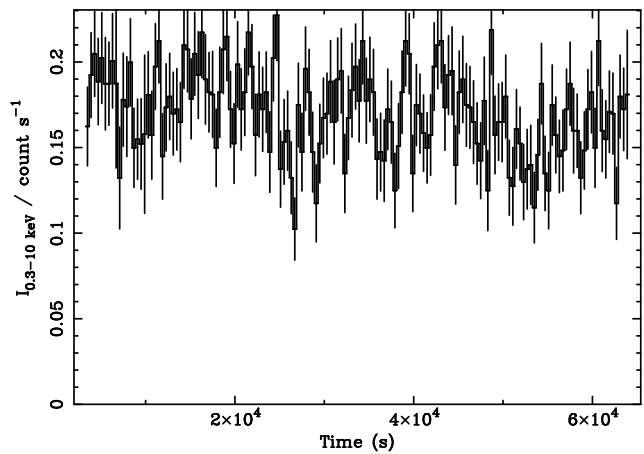}}
\caption{ Combined, background-subtracted pn+MOS lightcurve for XBo 144 in the 0.3--10 keV band. The lightcurve is binned to 400 s.}\label{lc} 
\end{figure}

The combined pn+MOS, background-subtracted, 0.3--10 keV lightcurve of XBo144 is shown in Fig.~\ref{lc}, with 400 s binning. The intensity appears to vary in the manner of low state LMXBs; however, the uncertainties are large, and the best fit line of constant intensity yields $\chi^2$/dof = 630/601, i.e. the lightcurve is consistent with being constant. The fractional r.m.s. variability is 6$\pm$4\% on time-scales longer than 100 s; this is consistent with a LMXB in the low state, but adds no extra strength to our argument. Without the definite variability exhibited by XBo 45 \citep{bsh08}, the case for XBo 144 being a black hole X-ray binary is somewhat weaker, and  rests on its association with the globular cluster Bo 144.

\section{Discussion}
\label{discuss}

The X-ray source associated with the M31 GC Bo 144, XBo 144, exhibits an emission spectrum that is characteristic of a LMXB in the low state: the 0.3--10 keV  emission is described by a pure power law, with $\Gamma$ = 1.48$\pm$0.04. \citet{glad07} showed that neutron star LMXBs  only exhibit this behaviour at 0.01--1000 keV luminosities $\la$10\%  of the Eddington limit; the 0.3--10 keV luminosity of XBo 144 exceeds 0.20 $L_{\rm Edd}$ for all known neutron stars (2.1 M$_{\odot}$ or less). 

Estimating the 0.01--1000 keV luminosity of XBo 144 is not as simple as extrapolating the power law to 1000 keV. Low state spectra are thought to be  caused by unsaturated inverse Compton scattering of cool photons on hot electrons in a corona, resulting in a power law  for energies lower than the electron temperature, and a Wien spectrum for higher energies \citep{sun80}.  The coronae in black hole X-ray binaries tend to have temperatures of $\sim$100--300 keV \citep{mr03}; hence we cannot determine the corona temperature with 0.3--10 keV data.  We obtained lower limit estimates  to the 0.01--1000 keV luminosity of  XBo 144 by measuring the flux of a power law with $\Gamma$=1.48 for energy ranges 0.3--100 keV and and 0.3--300 keV. We therefore expect the 0.01--1000 keV luminosity of XBo 144 to be  at least $\sim$3--7 times higher than the observed 0.3--10 keV luminosity, i.e. $>$0.6--1.4 $L_{\rm Edd}$ for a neutron star with mass 2.1 $M_{\odot}$.  Hence, we propose XBo 144 as a candidate GC black hole LMXB.

Without the strong time-variability observed in XBo 45, it is possible that XBo 144 is simply an active galactic nucleus (AGN), since $\Gamma$ $\sim$1.4 for typical AGN. Hence, associating the X-ray source with a GC is strong evidence that it is an X-ray binary.  We therefore calculated the probability of an AGN coinciding with a GC within the observed field of view. We employed the X-ray luminosity functions (XLFs) for AGN devised by \citet{mor03}. They created AGN XLFs in 1--2 keV and 2--10 keV bands; therefore we calculated the absorbed  1--2 and 2--10 keV fluxes of XBo 144 (1.08$\times$10$^{-13}$ and 5.69$\times$10$^{-13}$ erg cm$^{-2}$ s$^{-1}$, respectively). From the XLFs of Moretti et al. (2003), we expect 3.8$\pm$0.9$\times$10$^{-7}$ AGN per square arcsec to exhibit 1--2 keV fluxes as bright as XBo 144, and only 8$\pm$5$\times$10$^{-8}$ AGN per square arcsec to exhibit 2--10 keV fluxes as bright as XBo 144. \citet{K02} found the Chandra position of XBo 144 to be 2.2$''$ from the GC position. There are 20 GCs within the field of view of the XMM-Newton observation we are interested in; hence, the probability for a chance coincidence between a GC and an AGN as bright as Bo 144 in the 2--10 keV band  within 2.2$''$ is 2.4$\pm$1.5$\times$10$^{-5}$.

The 0.3--10 keV lightcurve of XBo 144 is consistent with a low state LMXB, but also is consistent with being constant.  We note that XBo 144 is $\sim$5 times fainter than XBo 45, and may be more severely limited by low photon counts.  There is a remarkable similarity between the two systems: they exhibit very similar emission spectra, they are both associated with GCs, and both appear to have been persistently bright over the $\sim$30 year period of observation. We infer that XBo 144 and XBo 45 are the same class of object: black hole LMXBs, likely produced by  tidal capture of a main sequence star.

We now consider the influence of metallicity on the probability of finding black hole LMXBs in GCs.  \citet{bel95} discovered  a novel mechanism to promote tidal capture  at higher metallicities, in addition to the previously suspected correlation between metallicity and initial mass function (IMF). They assume that for a fixed cluster density, the rate of tidal capture depends on the mass and radius of the capturing star. They find  that higher metallicity stars  have larger masses and radii; therefore, they expect the tidal capture rate to increase with metallicity. Furthermore, a metal-rich star will more easily fill its Roche lobe. They conclude that this effect alone could explain the observed ratio between frequencies of X-ray sources in metal-rich and metal-poor clusters, although there is no reason to exclude the effects of metallicity  on the IMF.    

\citet{fan08} have produced the most comprehensive survey of metallicities of M31 GCs yet made. They combine spectroscopically-derived metallicities for 295 GCs and GC candidates with colour-derived metallicities for 209 GCs and candidates. \citet{fan08} found a mean [Fe/H] of $-$1.29$\pm$0.03 for the combined sample, which is remarkably similar to the mean [Fe/H] for the Milky Way \citep[$-$1.30$\pm$0.05][ updated in 2003]{har96}. 
In particular, [Fe/H] = $-$0.6$\pm$0.2 for Bo 144, making it $\sim$5 times more metal rich than the mean for M31 or the Milky Way. Bo 45 also has a relatively high metallicity: [Fe/H] = $-$1.1$\pm$0.3, suggesting that it is $\sim$60\% more metal rich than the mean, with possible metallicity range 80--300\% of the M31 mean. The  high metallicity of Bo 144, and the inferred increase in tidal capture rate, support the plausibility of XBo 144 being  black hole LMXB.

\section{Conclusions}
\label{conc}

We have identified a second black hole candidate in a M31 GC; such a system would most likely be formed by tidal capture. 
\citet{kal04} found tidal capture to be the dominant mechanism for LMXB formation in collapsed core GCs, or even denser clusters such as M15. A survey of Galactic clusters \citep{har96} showed $\sim$20\% of clusters to be core collapsed; therefore, we  expect no more than 20\% of GC X-ray sources to harbour black holes. Traditional methods of identifying a black hole involve obtaining a mass function from the radial velocity curves of the donor, and are quite unsuitable for identifying black holes in GC LMXBs because the donor is near-impossible to identify. Our X-ray identification of black holes is not so hindered, but does identify only that subset of black hole LMXBs in the low state at an observed luminosity $\ga$3$\times$10$^{37}$ erg s$^{-1}$.  Applying our X-ray technique to other known, bright X-ray sources in globular clusters may reveal further candidate black hole systems.

\section*{Acknowlegments}
We thank the anonymous referee for their constructive comments. This work is based on observations with XMM-Newton, an
ESA science mission with instruments and contributions directly
funded by ESA member states and the US (NASA). Astronomy research at the Open University is funded by an STFC Rolling Grant.

\bibliographystyle{aa}
\bibliography{mnrasm31}
\label{lastpage}

\end{document}